\documentclass[11pt]{article}
\usepackage{amsfonts}
\usepackage{amsmath}
\usepackage{amssymb}
\usepackage{cite}
\usepackage{slashed}
\usepackage{epsfig}
\usepackage{graphicx}
\usepackage{array}
\usepackage{multirow}
\usepackage{color}
\usepackage[colorlinks=true,linkcolor=red,citecolor=blue]{hyperref}
\usepackage{url}
\begin{document}
\newcommand{\beq}{\begin{eqnarray}}
\newcommand{\eeq}{\end{eqnarray}}
\newcommand{\non}{\nonumber\\ }
\newcommand{\acp}{ {\cal A}_{CP} }
\newcommand{\psl}{ p \hspace{-1.8truemm}/ }
\newcommand{\nsl}{ n \hspace{-2.2truemm}/ }
\newcommand{\vsl}{ v \hspace{-2.2truemm}/ }
\newcommand{\epsl}{\epsilon \hspace{-1.8truemm}/\,  }
\newcommand{\tf}{\textbf}
\title{Study of $B_c \to DS$ Decays in PQCD Approach}
\author{Zhi-Tian Zou\footnote{zouzt@ytu.edu.cn },\,\,
Ying Li\footnote{liying@ytu.edu.cn},  \\
{\small \it Department of Physics, Yantai University, Yantai 264005,China}\\
\\
Xin Liu\footnote{liuxin@jsnu.edu.cn}\\
{\small \it School of Physics and Electronic Engineering, Jiangsu Normal University, Xuzhou 221116, China}
}
\maketitle
\begin{abstract}
Within the framework of perturbative QCD approach based on $k_T$ factorization, we studied 40 $B_c \to DS$ decay modes in the leading order and leading power, where ``S" stands for the light scalar meson. Under two different scenarios (S1 and S2) for the description of scalar mesons, we explored the branching fractions and related CP asymmetries. As a heavy meson consisting of two heavy quarks with different flavor, the light-cone distribution amplitude of $B_c$ meson has not been well defined, so the $\delta$-function is adopted. The contributions of emission diagrams are suppressed by the vector decay constants and CKM elements, the contributions of annihilation are dominant.  After the calculation, we found some branching fractions are in the range of  $[10^{-5}, 10^{-4}]$, which could be measured in the current LHCb experiment, and other decays with smaller branching fractions will be tested at the high-energy colliders in future. Furthermore, some decay modes have large CP asymmetries, but they are unmeasurable currently due to the small branching fractions.
\end{abstract}
\newpage
\section{Introduction}\label{sec:1}

The studies of weak decays of $B_c$ meson are of interest, since it is the only heavy meson consisting of two heavy quarks with different flavor. The Collider Detector at Fermilab (CDF) Collaboration reported the discovery of the $B_c$ ground state in $p\bar p$ collisions \cite{cdf}, which was further confirmed by the CDF and D0 Collaborations \cite{cdfd0} with more precise measurements. Currently, with high collision energy and high luminosity, Large Hadron Collider could collect about $10^9 $ events of $B_c$ meson each year \cite{hejb}. Based on such large samples, many weak decay modes of $B_c$ meson have been measured by LHCb collaboration \cite{lhcBC}.

In the quark model, the $B_c^+$ meson is the lowest-lying bound state of a bottom antiquark and a charm quark with $J^P=0^-$.  Since it carries flavor explicitly and can not annihilate into gluons, it is stable against the strong and electromagnetic annihilation processes and can only decay weakly, which provides a new window for studying the weak decay mechanism of heavy flavors. The another characteristic feature of the $B_c$ meson is that both quarks forming it are heavy and thus their weak decays give comparable contributions to the total decay rate. Therefore, the weak decay of $B_c$ meson can be categorized into into three classes: (i) the $b$ quark decays ($b\to c, u$) with the $c$ quark as a spectator, which can be used to precisely determinate the Cabibbo-Kobayashi-Maskawa (CKM) matrix elements $|V_{cb}|$ and $|V_{ub}|$, (ii) the $c$ quark decays ($c\to s,d$) with the $b$ quark as a spectator, which are suppressed by the phase space, but enhanced by the large CKM matrix element $|V_{cs}|$ or $|V_{cd}|$, and (iii) $b$ quark and $c$ quark co-annihilate, which are enhanced by $|V_{cb}/V_{ub}|^2 \sim 10^2$, in contrast to $B_u$ annihilation decays.  The estimations of the $B_c$ decay rates indicate that the $c$ quark decays give the dominant contribution ($\sim  70\%$) while the $b$ quark decays and weak annihilation contribute about $\sim  20\%$ and $\sim 10\%$, respectively. All in all, the $B_c$ meson provides very rich weak decay channels to study the perturbative and nonperturbative  QCD dynamics, annihilation mechanism in $B$ meson, to test the standard model, as well as to search for the signal of new physics \cite{yiyi}. In recent years, stimulated by both theoretical and experimental developments, many theoretical studies on the production and the semileptonic and nonleptonic decays of the $B_c$ meson have been explored by many groups based on Isgur-Scora-Grinstein-Wise (ISGW) quark model \cite{isgw}, the relativistic independent quark model \cite{riqm}, the QCD factorization \cite{bcqcdf1,bcqcdf2}, the light-front quark model \cite{bcqcdf2}, the SU(3) flavor symmetry \cite{su3}, the lattice gauge simulations \cite{lattice}, the sum rules \cite{sum}, the nonrelativistic QCD methods \cite{nrqcd}, and the perturbative QCD (PQCD) approach \cite{bcpqcd1,bcpqcd2,bcpqcd3}.

Specially, in refs.\cite{bcpqcd1,bcpqcd2,bcpqcd3}, the authors, including one of us (Zou), had investiaged the $B_c \to D^{(\star)}P(V)$ ($P$ and $V$ denoting the light pseudoscalar and vector meson) decays systematically within the PQCD approach \cite{kt1} that is based on the $k_T$ factorization. As known to all,  the $B_c$ meson is a nonrelativistic heavy quarkonium system, then the two quarks in the $B_c$ meson are both at rest and nonrelativistic. Since the charm quark in the final-state $D$ meson is almost at collinear state, a hard gluon is required to transfer large momentum to the spectator charm quark. So, the expansion based on $\alpha_s$ is reliable here. Moreover, we postulated a hierarchy $m_{B_c}\gg  m_{D^{(*)}}  \gg\Lambda_{QCD} $, the relation $m_{B_c}\gg m_{D^{(*)}} $ justifies the perturbative analysis of the $ B_c \to D^{(*)}  $ form factors at large recoil and the definition of light-cone $D^{(*)}$ meson wave functions. The relation $m_{D^{(*)}} \gg\Lambda_{QCD}$ justifies the power expansion in the parameter $\Lambda_{QCD}/m_{D^{(*)}}$.  The small ratio $\Lambda_{QCD}/m_B$ is viewed as being of higher power.  So, the factorization theorem is applicable to the $B_c$ system similar to the situation of the $B$ meson with a light quark. Utilizing the $k_T$ factorization instead of collinear factorization, this approach is free of endpoint singularity, so the diagrams including factorizable, nonfactorizable, as well as annihilation type diagrams are all calculable. In refs.\cite{bcpqcd1,bcpqcd2,bcpqcd3}, it showed us that some branching fractions are at the order of ${\cal O}(10^{-5})$, which is measurable in the running LHCb experiment. For completeness, in this work, we will extend previous studies to $B_c \to D^{(\star)}S$ decays where $S$ denotes a light scalar meson.

The light scalar mesons considered in this paper include the isosinglet $f_0(600)$($\sigma$), $f_0(980)$, $f_0(1370)$, $f_0(1500)$/$f_0(1710)$, the isodoublet $K_{0}^{*}(800)$($\kappa$) and $K_{0}^{*}(1430)$, and the isovector $a_0(980)$ and $a_0(1450)$  \cite{pdg}. In the literatures, the scalar mesons have been identified as ordinary $\bar{q}q$ states, four-quark states or meson-meson bound states or even those supplemented with a scalar glueball, however, a definite conclusion has not been obtained yet till now. In light of the mass spectrum of scalar mesons and the strong as well as electromagnetic decays, most of us accept that  the scalar mesons with the mass below 1 GeV constitute one nonet, while those near 1.5 GeV form another one \cite{nonet}. Moreover, the scalar meson states above 1 GeV can be identified as a conventional $q\bar q$  nonet with some possible glue content. However, the quark structure of the light scalar mesons below or near 1 GeV has been quite controversial, though they are widely perceived as primarily the four-quark bound states. In the literatures \cite{509,510}, according to the category that the light mesons belong to, two typical scenarios for describing the scalar mesons were propounded. The scenario-1 (S1) is the naive two-quark model: the nonet mesons below 1 GeV, such as $\kappa$, $a_0(980)$, $f_0(980)$, and $\sigma$, are treated as the lowest lying states, and accordingly these near 1.5 GeV, such as $a_0(1450)$, $K_0(1430)$, $f_0(1370/1500)$, are the first orbitally excited states. In scenario-2 (S2),  the nonet mesons near 1.5 GeV are viewed as the lowest lying states, while the mesons below 1 GeV may be the exotic states beyond the quark model such as four-quark bound states. We have to stress that although many experimental data indicates that the light scalar mesons, such as $f_0(980)$ and $a_0(980)$, are predominately four-quark states, however in practice it is very hard for us to make quantitative predictions based on the four-quark picture,  because both the decay constants and the distribution amplitudes of $S$ are beyond the conventional quark model. Hence, we shall discuss only the two-quark scenario for light scalar mesons in the current work.

For these considered $B_c\to D^{(*)} S$ with a emitted scalar meson, the factorizable emission amplitudes that are proportional to the matrix element $\langle S|(V \pm A) |0\rangle $ will vanish or be tiny, because the neutral scalar mesons cannot be produced through the $(V\pm A)$ current and the decay constants of the charged scalar mesons are suppressed by the fairly small difference between the two running current quark masses of the scalar meson. In order to obtain the precise and reliable predictions, it is necessary for us to go beyond the naive factorization and calculate the contributions from the nonfactorizable diagrams, as well as the annihilation diagrams. We also note that for the considered $B_c \to D^{(*)} S$ decays, the annihilation type diagrams will provide sizable contributions to the amplitudes and even dominate the amplitudes due to the enhancement from the large CKM matrix elements $V_{cb}$ and $V_{cs(d)}$.  It is worth mentioning that, the PQCD approach is an effective approach for calculating the nonfactorizable and annihilation diagrams, which can be confirmed by the precise predictions for the $B \to J/\psi D$ \cite{ann1} and $B^0 \to D^-_s K^+$ decays \cite{ann2}. So, for these considered decay channels, the predictions in PQCD approach are reliable.

The remainder of the paper is organized as follows. The framework of PQCD, as well as the distribution amplitudes and decay constants of the mesons, are given in Section.\ref{sec:2}. In Section.\ref{sec:3}, we shall present the formulae of each amplitudes for each diagram.  The numerical results and discussions will be given in Section \ref{sec:4}. We will summarize this work in the last section.
\section{Framework}\label{sec:2}
In this work,  we describe the meson's momentum by using the light-cone coordinate. In the rest frame of the $B_c$ meson, the momentum of $B_c$, scalar meson and $D$ meson,  up to the order of $r_D^2$, as given by
\begin{eqnarray}
P_{B_c} =\frac{M_{B_c}}{\sqrt{2}}(1,1,{\bf 0_T}),   P_{2} = \frac{M_{B_c}}{\sqrt{2}}(1-r_D^2,0,{\bf0_T}),  P_{3} =\frac{M_{B_c}}{\sqrt{2}}(r_D^2,1,{\bf0_T}),
\end{eqnarray}
where $r_D=m_D/m_{B_c}$. Note that the  terms involving $r_S^2$($r_S=m_S/m_{B_c}$) are neglected in this work.
\subsection{PQCD Approach}
It is known to us that in studying exclusive hadron decays the main theoretical uncertainties are from the calculations of matrix elements. The key point of the PQCD approach is keeping the intrinsic transverse momenta of the inner quarks, which is the so-called $k_T$ factorization \cite{kt1}. The additional energy scale induced  by the transverse momenta will lead to the double logarithms in the QCD radiative corrections. Using the resummation technique, the double logarithms can be absorbed in the sudakov form factor, which can suppress the long-distance contribution \cite{sudakov}. This sudakov factor practically makes PQCD approach applicable. Moreover, due to the radiative corrections of the weak vertex, another type of double logarithms $\alpha_s \ln^2x$, $x$ being the momentum fraction of the inner quark, actually exists while $x\to 0$; and therefore, these large corrections should be also resumed, called threshold resummation \cite{jet}. As a result, the endpoint singularity in traditional collinear factorization can be smeared by this threshold factor.

There are several typical scales in the $B_c$ decays. In general, the factorization hypothesis is adopted to deal with processes with multi scales. As we already know, the physics higher than the scale of the $W$ boson mass $(m_W)$ can be calculated perturbatively, and the Wilson coefficients at $m_W$ scale can be obtained. With the help of the renormalization group techniques, we can get the Wilson coefficients from the $m_W$ scale to the $b$ quark mass ($m_b$) scale. The hard part between $m_b$ scale and the factorization scale ($t$) can be calculated perturbatively in the PQCD approach. The physics lower than the $t$ scale belong to the soft dynamics, which is nonperturbative but universal, which can be parameterized into meson wave functions. The wave functions could be determined from experiments, or studied by  the nonperturbative QCD approaches, such as QCD sum rules and lattice QCD. Therefore, in PQCD approach, the decay amplitude  can be written as the convolution of the Wilson coefficients $C(t)$, the hard kernel $H(x_i,b_i,t)$, and the hadronic wave functions $\Phi_{B,D,S}(x_i,b_i)$ \cite{amp},
\begin{eqnarray}
\mathcal{A}\sim\int dx_1 dx_2 dx_3 b_1 db_1 b_2 db_2 b_3 db_3 \times \mathrm{Tr}[C(t)\Phi_B(x_1,b_1)\nonumber\\
\times\Phi_{S }(x_2,b_2)\Phi_{D }(x_3,b_3) H(x_i,,b_i,t)S_t(x_i)e^{-S(t)}.
\label{eq:amplitude}
\end{eqnarray}
In above, $\mathrm{Tr}$ denotes the trace over Dirac and color indices, the $x_i(i=1,2,3)$ are the momentum fractions of the ``light" quark in each meson, and the $b_i$ are the conjugate variables of $k_{Ti}$ of the valence quarks. The $S_t(x_i)$ and $e^{-S(t)}$ are the threshole resummation and the Sudakov form factor, respectively.

\subsection{Wave Functions of $B_c$ and $D$ Meson}
In PQCD approach, the universal nonperturbative wave functions are the most important inputs. Unlike $B_{u,d,s}$ mesons, our knowledge of the LCDAs for $B_c$ meson is quite poor till now (for a recent view, see \cite{Bell:2008er}). Although it has often been viewed as heavy quarkonium, we adopt the same form as the $B$ meson \cite{bcpqcd2,bcpqcd3},
\begin{eqnarray}
\Phi_{B_c}(x,b)=\frac{i}{\sqrt{6}}[(\makebox[-1.5pt][l]{/}P+m_{B_c})\gamma_5\phi_{B_c}(x,b)].
\end{eqnarray}

Given $m_{B_c}\approx m_b + m_c$, the light-cone distribution amplitude $\phi_{B_c}(x,b)$ can be written as \cite{Bell:2008er}
\begin{eqnarray}
\phi_{B_c}(x,b)=\frac{f_{B_c}}{2\sqrt{6}}\delta(x-m_c/m_{B_c})\exp\left[-\frac{1}{2}\omega^2b^2\right],
\end{eqnarray}
where the $m_c$ and $m_b$ are the mass of the charm quark and the beauty quark, respectively. $f_{B_c}$ is the decay constant of the $B_c$ meson. The introduced fator $\exp\left[-\frac{1}{2}\omega^2b^2\right]$ represents the $k_T$ dependence in the PQCD approach. It should be emphasized that this simple form is the two-particle nonrelativistic LCDAs at the leading order where both heavy valence quarks just share the total momentum of the $B_c$ mesons according to their masses. Since there are no enough experiment measurements to constrain the wave function and the distribution amplitude of the $B_c$ meson, the relativistic corrections and contributions from higher Fock states are not included in this work.

For the charmed $D^{(*)}$ mesons, following ref. \cite{Dwave}, we define the light cone distribution amplitudes as
\begin{eqnarray}
&&\langle D(p)|q_{\alpha}(z)\overline{c}_{\beta}(0)|0\rangle=\frac{i}{2\sqrt{6}}\int_0^1 dx e^{ixp\cdot z}[\gamma_5(\makebox[-1.5pt][l]{/}p +m_D)\phi_D(x,b)]_{\alpha,\beta},\label{DWF1}\\
&&\langle D^*(p)|q_{\alpha}(z)\overline{c}_{\beta}(0)|0\rangle=\frac{-1}{2\sqrt{6}}\int_0^1 dx e^{ixp\cdot z}[\makebox[-1.5pt][l]{/}\epsilon_L(\makebox[-1.5pt][l]{/}p + m_{D^*})\phi_{D^*}^L(x,b)]_{\alpha,\beta},\label{DWF2}
\end{eqnarray}
where the distribution amplitudes are
\begin{eqnarray}
\phi_D(x,b)=\phi_{D^*}^{L}(x,b)=\frac{1}{2\sqrt{6}}f_{D^{(*)}}6x(1-x)[1+C_D(1-2x)]\exp[-\frac{1}{2}\omega_D^2b^2],
\end{eqnarray}
with $\omega_D=0.15\pm0.5$ GeV. In this work, the high-twist distribution amplitudes are not included either, because they are suppressed by $\Lambda_{QCD}/{m_{D^{(*)}}}$. As for the parameters $C_D$, are fitted from the $B\to DP(V)$ and $B_s\to D_sP(V)$ decays \cite{Dwave,Dparameter}, are set to be $C_D=0.5\pm0.1$ and $C_{D_s}=0.4\pm0.1$, respectively.
\subsection{Physics of Light Scalar Mesons}
With the developments of experimental side, many scalar states have been discovered. Theoretically, as aforementioned, there are two different scenarios for describe the scalar mesons in the quark model. The scenario-1(S1) is the typical 2-quark model: the nonet mesons below 1 GeV, including $f_0(600)(\sigma)$, $f_0(980)$, $K^*_0(800)(\kappa)$ and $a_0(980)$, belong to the lowest lying states, and the ones near 1.5 GeV, including $f_0(1370)$, $f_0(1500)$/$f_0(1700)$, $K^*_0(1430)$ and $a_0(1450)$ are viewed as  the first excited states. In this scenario, the quark components of the light scalar mesons are given as
 \begin{eqnarray}
 \sigma=\frac{1}{\sqrt{2}}(u\bar u+d\bar d),\;\;f_0=s\bar s,\nonumber\\
 a_0^+=u\bar d,\;a_0^0=\frac{1}{\sqrt{2}}(u\bar u+d\bar d)\;,a_0^-=d\bar
 u, \nonumber\\
 \kappa^+=u\bar s,\;\kappa^0=d\bar s,\;\bar\kappa^0=s\bar
 d,\;\kappa^-=s\bar u.
 \end{eqnarray}
Here the $\sigma$ and $f_0(980)$ has the ideal mixing.  In fact, the observed $D_s \to f_0(980) \pi^{+}$ decay shows the probability of the $s\overline{s}$ component of $f_0(980)$, while $\Gamma(J/\psi \to f_0(980)\omega)\sim \Gamma(J/\psi \to f_0(980)\phi)$ indicated the existence of the non-strange components \cite{f0980-4-1,f0980-4-2}. Based on the data, in the 2-quark model, the $\sigma$ and $f_0(980)$ might be the mixing states as
 \begin{eqnarray}
 &&|f_0(980)\rangle=|s\bar s\rangle \cos\theta +|n\bar n\rangle
 \sin\theta,\nonumber\\
 &&|\sigma\rangle=-|s\bar s\rangle \sin\theta+|n\bar n\rangle
 \cos\theta,
 \end{eqnarray}
with $|n\bar n\rangle=\frac{1}{\sqrt{2}}(u\bar u+d\bar d)$ and $\theta$ is the mixing angle.  As for the mixing angle $\theta$, we can determine it by various experimental measurements \cite{value,upper}. Currently, by analyzing the present experimental data, two possible ranges of $25^{\circ}<\theta<40^{\circ}$ and $140^{\circ}<\theta<165^{\circ}$ \cite{angle} are preferred. Similarly, the $f_0(1370)$ and $f_0(1500)$ are the mixing states of $n\bar n$, $s\bar s$ and glueball. In this paper, according to refs \cite{heavymix}, we neglect the tiny contribution from scalar glueball \cite{Cheng:2015iaa}  and simplify the mixing form as
\begin{eqnarray}
&&f_0(1370)=0.78|n\bar n\rangle+0.51|s\bar s\rangle,\nonumber\\
&&f_0(1500)=-0.54|n\bar n\rangle+0.84|s\bar s\rangle.
\label{eq:f02}
\end{eqnarray}

In the scenario-2 (S2), the nonet mesons near 1.5GeV are viewed as the lowest lying states, while the mesons below 1 GeV may be viewed as four-quark bound states. Because of the difficulty when dealing with four-quark states, we only do the calculation about the heavier nonet in S2.

Now, we shall discuss the decay constants and the distribution amplitudes of the scalar mesons. The two decay constants of scalar mesons are defined
 \begin{eqnarray}
 \langle S(p)|\bar
 q_2\gamma_{\mu}q_1|0\rangle=f_Sp_{\mu}\;,\; \langle S|\bar q_2q_1|0\rangle=m_S\bar f_S.
 \end{eqnarray}
In term of the charge conjugate invariance, neutral scalar mesons cannot be produced by the vector current, so we obtain
 \begin{equation}
 f_{\sigma}=f_{f_0}=f_{a_0^0}=0.
 \end{equation}
For other scalar mesons, the vector decay constant $f_S$ and scalar one $\overline f_S$  are related by the equation of motion
\begin{eqnarray}
\overline{f}_S=\mu f_S,\;\;\;
\mu=\frac{m_S}{m_2(\mu)-m_1(\mu)},
\end{eqnarray}
where $m_S$ is the mass of the scalar meson, and $m_1$, $m_2$ are the running current quark masses. Inputs of the scalar mesons in our calculation, include the decay constants, running quark masses in this paragraph and the Gegenbauer moments in the following, quote from \cite{510}.

In the 2-quark model, the wave function of the scalar meson is given by
 \begin{eqnarray}
 \langle S(P_S)|q(0)_j\bar q(z)_l |0\rangle =\frac{-1}{\sqrt{2N_c}}\int^1_0dxe^{ixp\cdot z}\{\not P_S\phi_S(x)
 +m_S\phi^s_S(x)+m_S(\not{n}\not{v}-1)\phi^T_S(x)\}_{jl},\label{LCDA}
 \end{eqnarray}
with the lightlike vectors $n=(1,0,\textbf{0}_T)$ and $v=(0,1,\textbf{0}_T)$.  The twist-2 light-cone distribution amplitude (LCDA) $\Phi_S(x)$ and twist-3 LCDAs $\phi_S^s(x)$ and $\phi_S^{\sigma}$  satisfy the normalization conditions
 \begin{eqnarray}
 \int_0^1 dx\phi_S(x)&=&\frac{f_S}{2\sqrt{2N_c}},\nonumber\\
 \int_0^1 dx\phi_S^s(x)&=&\int_0^1
 dx\phi_S^{\sigma}(x)=\frac{\bar f_S}{2\sqrt{2N_c}}.
 \end{eqnarray}
The LCDAs can be expanded in Gegenbauer polynomials as the following
form:
 \begin{eqnarray}
 \phi_S(x)&=&\frac{f_S}{2\sqrt{2N_c}}6x(1-x)\bigg[1+\mu_s\sum_{m=1}^{\infty}B_m(\mu)C_m^{{3/2}}(2x-1)\bigg],\\
  \phi_S^s(x)&=&\frac{\bar f_S}{2\sqrt{2N_c}}[1+\sum_{m=1}^{\infty}a_m(\mu)C_m^{1/2}(2x-1)],\\
 \phi_S^T(x)&=&\frac{d}{dx}\frac{\phi_S^{\sigma}(x)}{6}=\frac{\bar f_S}{2\sqrt{2N_c}}\frac{d}{dx}\bigg\{
 x(1-x)[1+\sum_{m=1}^{\infty}b_m(\mu)C_m^{3/2}(2x-1)]\bigg\},
 \end{eqnarray}
where $B_m(\mu)$, $a_m(\mu)$ and $b_m(\mu)$ are the Gegenbauer moments, and $C_m^{(3/2)}$ and $C_m^{1/2}$ are the Gegenbauer polynomials. The explicit values of $B_m(\mu)$ can be found in \cite{510}. For the twist-3 LCDAs, we adopt the asymptotic form for simplicity, though the values of $b_m(\mu)$ and $a_m(\mu)$ have been explored \cite{Lu:2006fr}.

\section{Analytic Formulae}\label{sec:3}
For the considered decays, the weak effective Hamiltonian $\mathcal{H}_{eff}$ in the transition matrix elements can be written as  \cite{heff}
\begin{eqnarray}
\mathcal{H}_{eff}=\frac{G_{F}}{\sqrt{2}}\left[\sum_{q=u,c} V_{qb}^{*}V_{qX}\left[C_{1}(\mu)O_{1}^{q}(\mu)+C_{2}(\mu)O_{2}^{q}(\mu)\right] -V_{tb}^{*}V_{tX}\sum_{i=3}^{10}C_{i}(\mu)O_{i}(\mu)\right],
\end{eqnarray}
where the $V_{qb(X)}$ and $V_{tb(X)}$  ($X=d,s$) are the CKM matrix elements, and $C_i(i=1\sim10)$ are the Wilson coefficients at the scale $\mu$. The $O_{i}\,(i=1,...,10)$ are the so-called local four-quark operators:
\begin{itemize}
\item current-current (tree) operators
\begin{eqnarray}
O_{1}^{q}=(\bar{b}_{\alpha}q_{\beta})_{V-A}(\bar{q}_{\beta}X_{\alpha})_{V-A},\;\;\;O_{2}^{q}=(\bar{b}_{\alpha}q_{\alpha})
_{V-A}(\bar{q}_{\beta}X_{\beta})_{V-A},
\end{eqnarray}
\item QCD penguin operators
\begin{eqnarray}
&&O_{3}=(\bar{b}_{\alpha}X_{\alpha})_{V-A}\sum_{q^{\prime}}(\bar{q}^{\prime}_{\beta}q^{\prime}_{\beta})_{V-A},\;\;\;
O_{4}=(\bar{b}_{\alpha}X_{\beta})_{V-A}\sum_{q^{\prime}}(\bar{q}^{\prime}_{\beta}q^{\prime}_{\alpha})_{V-A},\\
&&O_{5}=(\bar{b}_{\alpha}X_{\alpha})_{V-A}\sum_{q^{\prime}}(\bar{q}^{\prime}_{\beta}q^{\prime}_{\beta})_{V+A},\;\;\;
O_{6}=(\bar{b}_{\alpha}X_{\beta})_{V-A}\sum_{q^{\prime}}(\bar{q}^{\prime}_{\beta}q^{\prime}_{\alpha})_{V+A},
\end{eqnarray}
\item
electro-weak penguin operators
\begin{eqnarray}
&&O_{7}=\frac{3}{2}(\bar{b}_{\alpha}X_{\alpha})_{V-A}\sum_{q^{\prime}}e_{q^{\prime}}(\bar{q}^{\prime}_{\beta}q^{\prime}_{\beta})_{V+A},
\;\;O_{8}=\frac{3}{2}(\bar{b}_{\alpha}X_{\beta})_{V-A}\sum_{q^{\prime}}e_{q^{\prime}}(\bar{q}^{\prime}_{\beta}q^{\prime}_{\alpha})_{V+A},\\
&&O_{9}=\frac{3}{2}(\bar{b}_{\alpha}X_{\alpha})_{V-A}\sum_{q^{\prime}}e_{q^{\prime}}(\bar{q}^{\prime}_{\beta}q^{\prime}_{\beta})_{V-A},\;\;
O_{10}=\frac{3}{2}(\bar{b}_{\alpha}X_{\beta})_{V-A}\sum_{q^{\prime}}e_{q^{\prime}}(\bar{q}^{\prime}_{\beta}q^{\prime}_{\alpha})_{V-A},
\end{eqnarray}
\end{itemize}
where $\alpha$ and $\beta$ are the color indices, and $q^{\prime}=(u,d,s,c,b)$ are the active quarks at the scale $m_{b}$. The $(V-A)$ and $(V+A)$ are the left-handed and right-handed currents and defined as $(\bar{b}_{\alpha}q_{\beta})_{V-A}=\bar{b}_{\alpha}\gamma_{\mu}(1-\gamma_{5})q_{\beta}$ and $(\bar{q}^{\prime}_{\beta}q^{\prime}_{\alpha})_{V+A}=\bar{q}^{\prime}_{\beta}\gamma_{\mu}(1+\gamma_{5})q^{\prime}_{\alpha}$, respectively. The combined Wilson coefficients $a_{i}$ used can be defined as \cite{prd58094009}:
\begin{eqnarray}
&&a_{1}=C_{2}+C_{1}/3,\;\;\;\;\;\;a_{2}=C_{1}+C_{2}/3,\nonumber\\
&&a_{i}=C_{i}+C_{i+1}/3,\,i=3,5,7,9,\nonumber\\
&&a_{j}=C_{j}+C_{j-1}/3, \,j=4,6,8,10.
\end{eqnarray}

\begin{figure}[!tbp]
\begin{center}
\vspace{-2cm}
\centerline{\epsfxsize=10 cm \epsffile{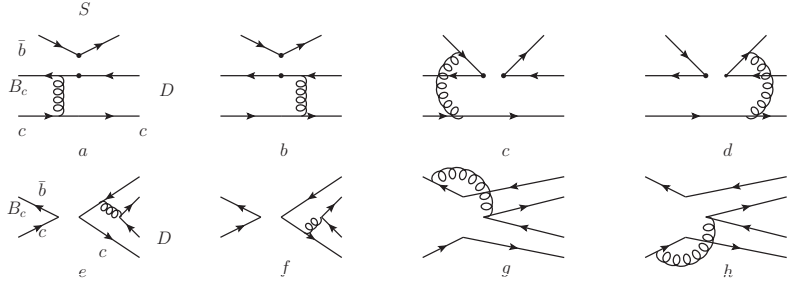}}
\vspace{-8.5cm}
\caption{Leading order Feynman diagrams contributing to the
$B_c\,\rightarrow\, D^{(*)}S$ decays in PQCD appraoch}
\label{fig:diagram1}
 \end{center}
\end{figure}

According to the effective Hamiltonian, we can draw the possible lowest order diagrams, as shown in Fig.\ref{fig:diagram1}, where the four diagrams in first line are the emission diagrams and four annihilation ones in second line. Now, we present the expressions of the hard kernels for all diagrams. After the perturbative calculation, when inserting different operators, the amplitudes for the factorizable emission diagrams in Figs.1a and 1b are given by
\begin{itemize}
\item $(V-A)(V-A)$
\begin{multline}
\mathcal{A}_{ef}^{LL}=8\pi C_f f_S m_{B_c}^4\int_0^1dx_1dx_3\int_0^{1/\Lambda}b_1db_1b_3db_3\phi_{B_c}(x_1,b_1)\phi_D(x_3,b_3)\\
\times\Big\{\Big[x_3(1-r_D^2)+r_b(r_D-2)-2r_Dx_3\Big] E_{ef}(t_a)h_a +\Big[(r_D-2)r_D\Big]E_{ef}(t_b)h_b\Big\},
\end{multline}
\item $(V-A)(V-A)$
\begin{eqnarray}
\mathcal{A}_{ef}^{LR}=\mathcal{A}_{ef}^{LL}
\end{eqnarray}
\item $(S-P)(S+P)$
\begin{multline}
\mathcal{A}_{ef}^{SP}=16\pi C_f \bar{f}_S r_S m_{B_c}^4\int_0^1dx_1dx_3\int_0^{1/\Lambda}b_1db_1b_3db_3\phi_{B_c}(x_1,b_1)\phi_D(x_3,b_3)\\
\times\Big\{\Big[r_b+r_D(1+x_3)-2\Big] E_{ef}(t_a)h_a
+\Big[2r_D(x_1-1)-x_1\Big]E_{ef}(t_b)h_b\Big\},
\end{multline}
\end{itemize}
In above formulas,  $r_b=m_b/m_{B_c}$  and $C_f=4/3$ is the group factor of $SU(3)_c$ for $B_c$ decays. The expressions of the scale $t$, Sudakov form factor $E$, and the hard functions $h$ can be found in the Appendix. The two diagrams (c) and (d) in Fig.\ref{fig:diagram1} are the nonfactorizable emission diagrams, whose contributions are as follows
\begin{itemize}
\item $(V-A)(V-A)$
 \begin{multline}
 \mathcal{M}_{enf}^{LL}=-16\sqrt{\frac{2}{3}}\pi C_f m_{B_c}^4\int_0^1dx_1dx_2dx_3\int_0^{1/\Lambda}b_1db_1b_2db_2\phi_{B_c}(x_1,b_1)\phi_S(x_2)\phi_D(x_3,b_1) \\
  \times\Big\{\Big[1-x_1-x_2+r_D(x_3-1)+r_D^2(x_1+2x_2-x_3)\Big] E_{enf}(t_c)h_c \\
 -\Big[1-x_1+x_2-x_3+r_D(x_3-1)+r_D^2(x_1-2x_2+x_3)\Big]   E_{enf}(t_d)h_d\Big\},
 \end{multline}
\item $(V-A)(V+A) $
\begin{multline}
 \mathcal{M}_{enf}^{LR}=16\sqrt{\frac{2}{3}}\pi C_f m_{B_c}^4\int_0^1dx_1dx_2dx_3\int_0^{1/\Lambda}b_1db_1b_2db_2\phi_{B_c}(x_1,b_1)\phi_D(x_3,b_1) \\
 \times\Big\{\Big[\phi_S^S(x_2)(x_1+x_2-1+r_D(x_1+x_2+x_3-2))  \\+\phi_S^{T}(x_2)(x_1+x_2-1+r_D(x_1+x_2-x_3))\Big]
\cdot E_{enf}(t_c)h_c \\
 -\Big[\phi_S^S(x_2)(x_1-x_2+r_D(x_1-x_2+x_3-1))\\-\phi_S^T(x_2)(x_1-x_2+r_D(x_1-x_2-x_3+1))\Big]
   \cdot E_{enf}(t_d)h_d\Big\},
 \end{multline}
\item $(S-P)(S+P) $
 \begin{multline}
 \mathcal{M}_{enf}^{SP}=-16\sqrt{\frac{2}{3}}\pi C_f m_{B_c}^4\int_0^1dx_1dx_2dx_3\int_0^{1/\Lambda}b_1db_1b_2db_2\phi_{B_c}(x_1,b_1)\phi_S(x_2)\phi_D(x_3,b_1) \\
  \times\Big\{\Big[2-x_1-x_2-x_3+r_D(x_3-1)+r_D^2(x_1+2x_2+x_3-2)\Big]
  \cdot E_{enf}(t_c)h_c \\
  -\Big[-x_1+x_2+r_D(x_3-1)+r_D^2(x_1-2x_2-x_3+2)\Big]
  \cdot E_{enf}(t_d)h_d\Big\}.
 \end{multline}
 \end{itemize}

As mentioned before, the annihilation type diagrams can be calculated reliably in PQCD approach.  For the factorizable annihilation diagrams (e) and (f), the expressions of hard kernels can be written as
\begin{itemize}
\item $(V-A)(V-A)$
\begin{multline}
\mathcal{A}_{af}^{LL}=8\pi C_f f_{B_c} m_{B_c}^4\int_0^1dx_2dx_3\int_0^{1/\Lambda}b_2db_2b_3db_3\phi_D(x_3,b_3) \\
 \times\Big\{\Big[\phi_S(x_2)(x_3-x_3r_D^2)+2\phi_S^S(x_2)r_S(x_3+1)r_D\Big]E_{af}(t_e)h_e \\
 -\Big[\phi_S(x_2)(x_2-(1+2x_2)r_D^2)+r_S(\phi_S^S(x_2)r_D(1+2x_2) \\
 -\phi_S^T(x_2)r_D(1-2r_Dx_2))\Big] E_{af}(t_f)h_f\Big\},
\end{multline}
\item $(V-A)(V+A)$
\begin{eqnarray}
\mathcal{A}_{af}^{LR}=\mathcal{A}_{af}^{LL}
\end{eqnarray}
\item $(S-P)(S+P) $
\begin{multline}
\mathcal{A}_{af}^{SP}=-16\pi C_f f_{B_c} m_{B_c}^4\int_0^1dx_2dx_3\int_0^{1/\Lambda}b_2db_2b_3db_3\phi_D(x_3,b_3) \\
 \times\Big\{\Big[\phi_S(x_2)r_Dx_3+2\phi_S^S(x_2)r_S\Big]E_{af}(t_e)h_e
 +\Big[\phi_S(x_2)r_D+r_Sx_2(\phi_S^S(x_2)-\phi_S^T(x_2))\Big]  E_{af}(t_f)h_f\Big\}.
\end{multline}
\end{itemize}

As for  nonfactorizable annihilation diagrams (g) and (h), the amplitudes  are as follows.
\begin{itemize}
\item $(V-A)(V-A)$
\begin{multline}
\mathcal{M}_{anf}^{LL}=-16\sqrt{\frac{2}{3}}\pi C_f m_{B_c}^4\int_0^1dx_1dx_2dx_3\int_0^{1/\Lambda}b_1db_1b_2db_2 \phi_B(x_1,b_1)\phi_D(x_3,b_2) \\
 \times\Big\{\Big[\phi_S(x_2)(1-x_1-x_2-r_b+r_D^2(x_1+2x_2-x_3))-r_Dr_S(\phi_S^S(x_2)(x_1+x_2+x_3-2) \\
 +\phi_S^T(x_2)(x_1+x_2-x_3))\Big]E_{anf}(t_g)h_g \\
+\Big[\phi_S(x_2)(r_D+x_3)+r_Dr_S(\phi_S^S(x_2)(x_2+x_3-x_1) \\
+\phi_S^T(x_2)(x_1-x_2+x_3))\Big]E_{anf}(t_h)h_h\Big\},
\end{multline}
\item $(V-A)(V+A)$
\begin{multline}
\mathcal{M}_{anf}^{LR}=-16\sqrt{\frac{2}{3}}\pi C_f m_{B_c}^4\int_0^1dx_1dx_2dx_3\int_0^{1/\Lambda}b_1db_1b_2db_2 \phi_B(x_1,b_1)\phi_D(x_3,b_2) \\
\times\Big\{\Big[\phi_S(x_2)r_D(x_3-2)-r_S(\phi_S^S(x_2)+\phi_S^T(x_2))(x_1+x_2-2)\Big] E_{anf}(t_g)h_g \\
+\Big[\phi_S(x_2)r_D(r_D-x_3)-r_S(\phi_S^S(x_2)+\phi_S^T(x_2))(r_D+x_1-x_2)\Big] E_{anf}(t_h)h_h\Big\},
\end{multline}
\item $(S-P)(S+P)$
\begin{multline}
\mathcal{M}_{anf}^{SP} =16\sqrt{\frac{2}{3}}\pi C_f m_{B_c}^4\int_0^1dx_1dx_2dx_3\int_0^{1/\Lambda}b_1db_1b_2db_2 \phi_B(x_1,b_1)\phi_D(x_3,b_2) \\
\times\Big\{\Big[\phi_S(x_2)(1-r_b-x_3)-r_Dr_S(\phi_S^S(x_2)(x_1+x_2+x_3-2) \\
-\phi_S^T(x_2)(x_1+x_2-x_3))\Big]E_{anf}(t_g)h_g \\
-\Big[\phi_S(x_2)(x_1-x_2-r_D-r_D^2(x_1-2x_2+x_3))+r_Dr_S(\phi_S^S(x_2)(x_1-x_2+x_3) \\
-\phi_S^T(x_2)(x_2+x_3-x_1))\Big]E_{anf}(t_h)h_h\Big\}.
\end{multline}
\end{itemize}

Similarly, the formulas for the $B_c \to D^* S$ decays are as follows:
\begin{multline}
\mathcal{A}_{ef}^{*LL}= -8\pi C_f f_S m_{B_c}^4\int_0^1dx_1dx_3\int_0^{1/\Lambda}b_1db_1b_3db_3
\phi_{B_c}(x_1,b_1)\phi^{L}_{D^*}(x_3,b_3) \\
 \times \Big \{\Big[r_b(r_D-2)-2r_Dx_3+x_3(1-r_D^2)\Big]E_{ef}(t_a)h_a
 -r_D^2E_{ef}(t_b)h_b\Big\},
\end{multline}
\begin{eqnarray}
\mathcal{A}_{ef}^{*LR}=\mathcal{A}_{ef}^{*LL},
\end{eqnarray}
\begin{multline}
\mathcal{A}_{ef}^{*SP}=-16\pi C_f \bar{f}_S r_S m_{B_c}^4\int_0^1dx_1dx_3\int_0^{1/\Lambda}b_1db_1b_3db_3\phi_{B_c}(x_1,b_1)\phi_{D^*}^L(x_3,b_3) \\
\times \Big\{\Big[2-r_b+r_D(1-x_3)\Big] E_{ef}(t_a)h_a  +x_1E_{ef}(t_b)h_b\Big\},
\end{multline}
\begin{multline}
 \mathcal{M}_{enf}^{*LL}=16\sqrt{\frac{2}{3}}\pi C_f m_{B_c}^4\int_0^1dx_1dx_2dx_3\int_0^{1/\Lambda}b_1db_1b_2db_2\phi_{B_c}(x_1,b_1)\phi_S(x_2)\phi_{D^*}^L(x_3,b_1) \\
  \times \Big\{\Big[1-x_1-x_2+r_D(1-x_3)+r_D^2(x_1+2x_2+x_3-2)\Big]
  \cdot E_{enf}(t_c)h_c \\
 -\Big[1-x_1+x_2-x_3+r_D(x_3-1)+r_D^2(x_1-2x_2+x_3)\Big]
  \cdot E_{enf}(t_d)h_d\Big\},
 \end{multline}
 \begin{multline}
 \mathcal{M}_{enf}^{*LR}=-16\sqrt{\frac{2}{3}}\pi C_f m_{B_c}^4\int_0^1dx_1dx_2dx_3\int_0^{1/\Lambda}b_1db_1b_2db_2\phi_{B_c}(x_1,b_1)\phi_{D^*}^L(x_3,b_1) \\
 \times\Big\{\Big[\phi_S^S(x_2)((r_D-1)(x_1+x_2)+r_Dx_3+1) \\
 +\phi_S^{T}(x_2)(1-x_1-x_2+r_D(x_1+x_2+x_3-2))\Big]
 \cdot E_{enf}(t_c)h_c \\
 -\Big[\phi_S^S(x_2)(x_2-x_1+r_D(x_1-x_2-x_3+1))  \\
 -\phi_S^T(x_2)(x_2-x_1+r_D(x_1-x_2+x_3-1))\Big]
 \cdot E_{enf}(t_d)h_d\Big\},
 \end{multline}
 \begin{multline}
 \mathcal{M}_{enf}^{*SP}=16\sqrt{\frac{2}{3}}\pi C_f m_{B_c}^4\int_0^1dx_1dx_2dx_3\int_0^{1/\Lambda}b_1db_1b_2db_2\phi_{B_c}(x_1,b_1)\phi_S(x_2)\phi_{D^*}^L(x_3,b_1) \\
  \times \Big\{\Big[r_D^2(x_1+2x_2+x_3-2)+r_D(x_3-1)+2-x_1-x_2-x_3\Big]
 E_{enf}(t_c)h_c \\
  -\Big[r_D^2(x_1-2x_2+x_3)+r_D(1-x_3)-x_1+x_2\Big]
 E_{enf}(t_d)h_d\Big\},
 \end{multline}
\begin{multline}
\mathcal{A}_{af}^{*LL}=-8\pi C_f f_{B_c} m_{B_c}^4\int_0^1dx_2dx_3\int_0^{1/\Lambda}b_2db_2b_3db_3\phi_{D^*}^L(x_3,b_3) \\
\times\Big\{\Big[\phi_S(x_2)(1-r_D^2)x_3+2\phi_S^S(x_2)r_Sr_D(x_3-1)\Big]E_{af}(t_e)h_e \\
 -\Big[\phi_S(x_2)(r_D^2(1-2x_2)+x_2)+r_Dr_s(\phi_S^S(x_2)-\phi_S^T(x_2))\Big]  E_{af}(t_f)h_f\Big\},
\end{multline}
\begin{eqnarray}
\mathcal{A}_{af}^{*LR}&=&\mathcal{A}_{af}^{*LL},
\end{eqnarray}
\begin{multline}
\mathcal{A}_{af}^{*SP}=16\pi C_f f_{B_c} m_{B_c}^4\int_0^1dx_2dx_3\int_0^{1/\Lambda}b_2db_2b_3db_3\phi_{D^*}^L(x_3,b_3) \\
 \times \Big\{\Big[\phi_S(x_2)r_Dx_3-2\phi_S^S(x_2)r_S\Big]E_{af}(t_e)h_e \\
 -\Big[\phi_S(x_2)r_D-r_Sx_2(\phi_S^T(x_2)-\phi_S^S(x_2))\Big]E_{af}(t_f)h_f\Big\},
\end{multline}
\begin{multline}
\mathcal{M}_{anf}^{*LL}=16\sqrt{\frac{2}{3}}\pi C_f m_{B_c}^4\int_0^1dx_1dx_2dx_3\int_0^{1/\Lambda}b_1db_1b_2db_2 \phi_B(x_1,b_1)\phi_{D^*}^L(x_3,b_2) \\
 \times\Big\{\Big[\phi_S(x_2)(1-r_b-x_1-x_2+r_D^2(x_1+2x_2+x_3-2))
 -r_Dr_S(\phi_S^S(x_2)(x_1+x_2-x_3)\\
 +\phi_S^T(x_2)(x_1+x_2+x_3-2))\Big] \cdot E_{anf}(t_g)h_g \\
 +\Big[\phi_S(x_2)(r_D+(1-2r_D^2)x_3)+r_Dr_S(\phi_S^S(x_2)(x_1-x_2+x_3) \\+\phi_S^T(x_2)(x_2+x_3-x_1))\Big]
\cdot E_{anf}(t_h)h_h \Big\},
\end{multline}
\begin{multline}
\mathcal{M}_{anf}^{*LR}=-16\sqrt{\frac{2}{3}}\pi C_f m_{B_c}^4\int_0^1dx_1dx_2dx_3\int_0^{1/\Lambda}b_1db_1b_2db_2 \phi_B(x_1,b_1)\phi_{D^*}^L(x_3,b_2) \\
 \times\Big\{\Big[\phi_S(x_2)r_D(x_3-2)-r_S(\phi_S^S(x_2)+\phi_S^T(x_2))(x_1+x_2-2)\Big]
    E_{anf}(t_g)h_g  \\
 +\Big[\phi_S(x_2)r_D(r_D-x_3)-r_S(\phi_S^S(x_2)+\phi_S^T(x_2))(r_D+x_1-x_2)\Big]
 E_{anf}(t_h)h_h \Big\},
\end{multline}
\begin{multline}
\mathcal{M}_{*anf}^{SP}=-16\sqrt{\frac{2}{3}}\pi C_f m_{B_c}^4\int_0^1dx_1dx_2dx_3\int_0^{1/\Lambda}b_1db_1b_2db_2 \phi_B(x_1,b_1)\phi_{D^*}^L(x_3,b_2) \\
 \times\Big\{\Big[\phi_S(x_2)(r_b+(1-2r_D^2)(x_3-1))\\+r_Dr_S(\phi_S^T(x_2)(x_1+x_2+x_3-2)
  -\phi_S^S(x_2)(x_1+x_2-x_3))\Big]E_{anf}(t_g)h_g  \\
 +\Big[\phi_S(x_2)(x_1-x_2-r_D-r_D^2(x_1-2x_2-x_3))\\+r_Dr_S(\phi_S^S(x_2)(x_1-x_2+x_3)
-\phi_S^T(x_2)(x_1-x_2-x_3))\Big]E_{anf}(t_h)h_h \Big\}.
\end{multline}

Note that the total decay amplitudes containing the Wilson coefficients and the CKM elements are the same as the
$B_c \to D^{(*)} P$ decays with $P$ denoting a pseudoscalar meson, which can be found in the ref.\cite{bcpqcd2}, because the topologies of these two type decays are identical. As an example, we show the total decay amplitude of $ B_c\rightarrow D^{+}K_{0}^{*0}$ as
\begin{eqnarray}
\mathcal{A}(B_c\rightarrow D^{+}K_{0}^{*0})&=&\frac{G_{F}}{\sqrt{2}}\Big \{V_{cb}^{*}V_{cs}(\mathcal{M}_{af}^{LL}a_{1}+\mathcal{M}_{anf}^{LL}C_{1}) \nonumber\\
&& -V_{tb}^{*}V_{ts}[\mathcal{M}_{enf}^{LL}(C_{3}-C_{9}/2)+\mathcal{M}_{enf}^{LR}(C_{5}-C_{7}/2)+\mathcal{M}_{af}^{LL}(a_{4}+a_{10}) \nonumber\\
&& +\mathcal{M}_{af}^{SP}(a_6+a_{8})+\mathcal{M}_{anf}^{LL}(C_3+C_9)+\mathcal{M}_{anf}^{LR}(C_5+C_7)]\Big\}.
\end{eqnarray}
The decay width of  $B_c \to D^{(*)}S$  is given by
\begin{eqnarray}
\Gamma(B_c \to D^{(*)}S)=\frac{p}{8\pi
m_{B_c}^{2}}|\mathcal {A}(B_c \to D^{(*)}S)|^{2},
\end{eqnarray}
where the momentum of the final state particle is
\begin{eqnarray}
p=\frac{1}{2m_{B_c}}\sqrt{\left[m_{B_c}^{2}-(m_{D}+m_{S})^{2}\right]\left[m_{B_c}^{2}-(m_{D}-m_{S})^{2}\right]} .
\end{eqnarray}

\section{Numerical Results and Discussions}\label{sec:4}
In this section, we shall firstly list the other  input parameters we used in  the numerical calculations, such as the masses and life times of mesons, the CKM matrix elements, as following \cite{pdg}
\begin{eqnarray}
 &&\Lambda_{\overline{MS}}^{f=4}=0.25\pm0.05 \mathrm{GeV},\;\;m_{B_{c}}=6.28\mathrm{GeV},\;\;m_b=4.8\mathrm{GeV},\nonumber\\
 &&m_{D_{(s)}}=1.87/1.97\mathrm{GeV},\;\;m_{D_{(s)}^*}=2.01/2.11\mathrm{GeV},\nonumber\\
  &&\tau_{B_{c}}=0.46 ps,\;\; \gamma=(69^{+10}_{-11})^{\circ},\;\;\lambda=0.225\pm 0.001,\nonumber\\
  &&V_{cb}=0.041_{-0.001}^{+0.001},\;\;V_{us}=0.225 \pm0.001 , \nonumber\\
  &&V_{td}=0.009_{-0.001}^{+0.001},\;\;V_{ts}=-0.040_{-0.001}^{+0.001}\;\;.
  \label{eq:parameter}
 \end{eqnarray}
 
Within the above parameters, we calculated the CP-averaged branching fractions and the direct CP asymmetries of all 40 $B_c\to D^{(*)} S$ decays and summarized the results in the Tables. \ref{T1}-\ref{T4}. We acknowledged that there are many uncertainties in our work. In tables, we mainly estimated three kinds of errors caused by the corresponding parameters. The first uncertainties come from the nonperturbative parameters, such as the decay constants and the distribution amplitudes. The second errors are from the high order corrections. Since the next-leading order correction have not yet been finished, we shall vary the range ($0.75 t\to 1.25t$) of the hard scale $t$ to estimate this kind uncertainty. This strategy has been widely used in the studies of $B$ meson decays. The last errors arise from the uncertainties of the CKM matrix elements. Unlike the $B_q$ mesons, $B_c$ decays are dominated by the factorizable annihilations, the amplitudes of which are in proportional to the decay constant of $B_c$, so the branching fractions are not sensitive to the the distribution amplitude of $B_c$ meson (less than $10\%$)\cite{bcpqcd3}. Therefore, we have not included the uncertainties taken by the $B_c$ wave function. We also emphasized that the next-leading power corrections will take large uncertainties, however this kind of study is beyond the scope of the current work, and we left them as our future work. For convenience, in above tables, we mark the dominant contributions of each decay mode  by the symbols ``C"(color-suppressed tree contributions), ``A"($W$ annihilation type contributions), and ``P'' (penguin contributions).
\begin{table}[!htb]
\centering
\caption{The CP-averaged branching fractions and CP asymmetries of $B_{c} \to D S(a_0(980),\kappa, \sigma, f_0(980))$ decays calculated in the PQCD approach in S1.}
\begin{tabular}[t]{l!{\;\;\;\;}c!{\;\;}c!{\;\;}c}
\hline\hline
\multirow{2}{*}{Decay Modes} & \multirow{2}{*}{Class} & \multirow{2}{*}{BFs($10^{-6}$)} &\multirow{2}{*}{$A_{CP}$($\%$)}\\
  &&\\
 \hline
 \vspace{0.05cm}
$B_c\to D^{+}a_{0}^0(980)$&A&$3.44_{-0.75-1.69-0.32}^{+0.83+1.73+0.34}$&$36.0_{-4.5-8.2-1.4}^{+5.0+7.0+0.5}$\\
\vspace{0.05cm}
$B_{c}\to D^{0}a_0^+(980)$&A&$6.78_{-1.53-3.44-0.49}^{+1.71+3.33+0.52}$&$1.88_{-2.09-3.93-0.79}^{+1.78+0.90+0.47}$\\
\vspace{0.05cm}
$B_c\to D^{+}\kappa^0_0(800)$ &A&$107_{-28-55-4}^{+32+65+3}$&$0.0$\\
\vspace{0.05cm}
$B_c \to D^0 \kappa^+(800)$&A&$93.6_{-24.7-50.7-3.0}^{+26.7+56.5+3.0}$&$0.91_{-0.21-0.25-0.09}^{+0.23+0.30+0.06}$\\
\vspace{0.05cm}
$B_c\to D^{+}\sigma(f_n)$& A & $3.39_{-0.71-1.30-0.10}^{+0.77+2.21+0.09}$&$-36.4_{-3.8-5.3-2.6}^{+3.4+6.4+3.0}$\\
\vspace{0.05cm}
$B_c\to D^{+}\sigma (f_s)$ & P & $0.02_{-0.01-0.01-0.00}^{+0.01+0.01+0.00}$&$0.0$\\
\vspace{0.05cm}
$B_c\to D^{+}f_0(980) (f_n)$ &A&$4.60_{-0.93-1.70-0.17}^{+1.04+2.19+0.16}$&$-35.3_{-3.7-4.1-2.2}^{+3.3+6.4+2.6}$\\
\vspace{0.05cm}
$B_c\to D^{+}f_0(980)(f_s)$&P&$0.02_{-0.01-0.01-0.00}^{+0.01+0.01+0.00}$&$0.0$ \\
\vspace{0.05cm}
$B_c\to D_s a_0^0(980)$&C&$0.03_{-0.01-0.01-0.00}^{+0.01+0.01+0.00}$&$-2.28_{-0.26-1.15-0.12}^{+0.26+0.64+0.16}$\\
\vspace{0.05cm}
$B_c\to D_s \sigma(f_n)$&P&$0.92_{-0.24-0.30-0.02}^{+0.27+0.46+0.03}$&$1.12_{-0.26-0.26-0.12}^{+0.32+0.34+0.10}$\\
\vspace{0.05cm}
$B_c\to D_s \sigma(f_s)$& A &$136_{-32-55-5}^{+32+94+4}$&$0.0$\\
\vspace{0.05cm}
$B_c\to D_{s}f_0(980)(f_n)$ &P&$0.92_{-0.24-0.26-0.02}^{+0.27+0.49+0.03}$&$1.12_{-0.26-0.26-0.12}^{+0.32+0.34+0.10}$\\
\vspace{0.05cm}
$B_c\to D_{s}f_{0}(980)(f_s)$&A&$188_{-39-61-6}^{+42+69+6}$&$0.0$\\
\vspace{0.05cm}
$B_{c}\to D_s \bar{\kappa}_0^{0}(800)$&A&$6.80_{-1.74-3.78-0.39}^{+1.70+4.45+0.40}$&$-9.26_{-1.74-3.09-0.06}^{+1.65+2.10+0.06}$\\
 \hline\hline
\end{tabular}\label{T1}
\end{table}
 \begin{table}[!ht]
 \begin{center}
\caption{The CP-averaged branching fractions and the CP asymmetries of $B_c \to D S$ ($a_0(1450), K_0^*(1430), f_0(1370)$, and $f_0(1500)$) calculated in the PQCD approach in S1 and S2, respectively.}
\begin{tabular}[t]{l!{\;}c!{\;}c!{\;}c!{\;}c}
\hline\hline
  \multirow{2}{*}{Decay Modes} & \multirow{2}{*}{Class} & \multirow{2}{*}{BFs($10^{-6}$)}&\multirow{2}{*}{$A_{CP}(\%)$}&\multirow{2}{*}{Scenario}\\
  &&&&\\
 \hline
 \vspace{0.05cm}
$B_c\to D^{+}a_{0}^0(1450)$&A&$5.30_{-1.72-1.20-0.03}^{+1.90+0.56+0.00}$&$-27.7_{-10.6-3.6-3.5}^{+11.4+2.5+3.7}$&$S_1$\\
\vspace{0.05cm}
&&$14.7_{-4.2-5.2-0.5}^{+4.7+4.7+0.8}$&$13.6_{-4.1-2.4-2.0}^{+1.9+1.4+0.3}$&$S_2$\\
\vspace{0.05cm}
$B_{c}\to D^{0}a_{0}^+(1450)$ &A&$9.89_{-3.26-2.84-0.32}^{+3.27+1.19+0.27}$&$-7.89_{-4.62-1.52-0.92}^{+3.78+0.00+1.10}$&$S_1$\\
\vspace{0.05cm}
&&$27.1_{-8.2-9.6-1.4}^{+8.9+9.9+1.5}$&$1.91_{-1.12-2.86-0.42}^{+0.86+2.00+0.29}$&$S_2$\\
\vspace{0.05cm}
$B_c\to D^{+}f_0(1370)$ &A&$1.22_{-0.37-0.38-0.17}^{+0.85+0.14+0.19}$&$54.3_{-33.1-1.3-1.8}^{+30.1+27.3+0.4}$&$S_1$\\
\vspace{0.05cm}
&&$7.01_{-2.06-2.67-0.46}^{+2.36+1.97+0.49}$&$-20.4_{-3.9-4.7-0.6}^{+3.8+1.6+1.1}$&$S_2$\\
\vspace{0.05cm}
$B_c\to D^{+}f_0(1500)$&A&$0.94_{-0.30-0.10-0.09}^{+0.40+0.14+0.11}$&$43.7_{-25.0-15.6-1.2}^{+28.1+19.0+0.4}$&$S_1$\\
\vspace{0.05cm}
&&$3.60_{-1.11-1.24-0.11}^{+1.28+1.48+0.13}$&$-22.8_{-5.4-3.0-0.9}^{+5.3+2.0+1.3}$&$S_2$\\
\vspace{0.05cm}
$B_c\to D^{+}K_0^{*0}(1430)$ &A&$191_{-43-49-6}^{+46+26+5}$&0.0&$S_1$\\
\vspace{0.05cm}
&&$481_{-166-216-13}^{+175+170+14}$&0.0&$S_2$\\
\vspace{0.05cm}
$B_{c}\to D^{0}K_{0}^{*+}(1430)$&A&$193_{-40-40-6}^{+50+29+6}$ &$1.10_{-0.15-0.18-0.11}^{+0.17+0.45+0.09}$&$S_1$\\
\vspace{0.05cm}
&&$458_{--166-176-13}^{+175+166+13}$&$0.24_{-0.11-0.09-0.02}^{+0.12+0.06+0.02}$&$S_2$\\
\vspace{0.05cm}
$B_{c}\to D_{s}a_0^{0}(1450)$ &C&$0.05_{-0.01-0.01-0.00}^{+0.02+0.02+0.00}$&$-1.94_{-0.45-1.07-0.10}^{+0.28+1.11+0.15}$&$S_1$\\
\vspace{0.05cm}
&&$0.02_{-0.01-0.01-0.00}^{+0.01+0.01+0.00}$&$0.50_{-0.36-4.47-0.58}^{+0.80+0.86+0.03}$&$S_2$\\
\vspace{0.05cm}
$B_{c}\to D_s f_0(1370)$&A&$22.8_{-9.3-7.7-1.2}^{+16.7+2.3+1.1}$&$-3.68_{-3.19-2.40-0.23}^{+2.53+0.16+0.30}$&$S_1$\\
\vspace{0.05cm}
&&$144_{-43-55-5}^{+46+42+5}$&$1.12_{-0.21-0.10-0.10}^{+0.21+0.15+0.08}$&$S_2$\\
\vspace{0.05cm}
$B_c\to D_s f_0(1500)$&A&$113_{-46-36-3}^{+67+16+4}$&$0.90_{-0.59-0.05-0.08}^{+0.83+0.51+0.07}$&$S_1$\\
\vspace{0.05cm}
&&$209_{-62-24-9}^{+66+31+8}$&$-0.66_{-0.09-0.06-0.05}^{+0.10+0.06+0.06}$&$S_2$\\
\vspace{0.05cm}
$B_{c}\to D_s \bar{K}_0^{*0}(1430)$&A&$9.17_{-2.25-0.81-0.53}^{+2.40+1.81+0.55}$&$7.18_{-096-1.91-0.05}^{+0.91+4.79+0.04}$&$S_1$\\
\vspace{0.05cm}
&&$27.9_{-9.2-9.2-1.6}^{+10.8+13.0+1.7}$&$-3.14_{-1.68-1.37-0.02}^{+1.06+1.36+0.02}$&$S_2$\\
\hline\hline
\end{tabular}\label{T2}
 \end{center}
\end{table}
  \begin{table}[!ht]
\centering
 \caption{The CP-averaged branching fractions and CP asymmetries of $B_{c} \to D^* S(a_0,\kappa, \sigma, f_0)$ decays calculated in the PQCD approach in S1.}
\begin{tabular}[t]{l!{\;\;\;\;}c!{\;\;}c!{\;\;}c}
\hline\hline
  \multirow{2}{*}{Decay Modes} & \multirow{2}{*}{Class} & \multirow{2}{*}{BFs($10^{-6}$)}&\multirow{2}{*}{$A_{CP}(\%)$} \\
  &&&\\
 \hline
 \vspace{0.05cm}
$B_c\to D^{*+}a_{0}^0(980)$&A&$3.38_{-0.59-1.34-0.20}^{+0.60+0.66+0.19}$&$-69.8_{-6.0-7.9-3.5}^{+6.5+4.5+4.9}$\\
\vspace{0.05cm}
$B_{c}\to D^{*0}a_0^+(980)$&A&$5.42_{-0.89-2.11-0.29}^{+0.90+1.51+0.29}$&$-28.4_{-3.6-4.3-1.7}^{+3.7+1.0+2.3}$\\
\vspace{0.05cm}
$B_c\to D^{*+}\sigma(f_n)$ &A&$1.65_{-0.25-0.59-0.06}^{+0.29+0.39+0.06}$&$71.6_{-4.1-5.2-6.5}^{+3.8+5.6+5.5}$\\
\vspace{0.05cm}
$B_c\to D^{*+}\sigma(f_s)$&P& $0.04_{-0.01-0.01-0.00}^{+0.01+0.01+0.00}$&0.0\\
\vspace{0.05cm}
$B_c\to D^{*+}f_0(980)(f_n)$ &A & $2.94_{-0.38-0.98-0.14}^{+0.40+0.57+0.13}$&$58.7_{-4.1-3.2-4.8}^{+3.8+6.0+3.9}$\\
\vspace{0.05cm}
$B_c\to D^{*+}f_0(980)(f_s)$ &P&$0.04_{-0.01-0.01-0.00}^{+0.01+0.01+0.00}$&0.0\\
\vspace{0.05cm}
$B_{c}\to D^{*+}\kappa^{0}(800)$&A&$54.8_{-9.1-23.3-1.6}^{+9.3+28.8+1.8}$&$0.0$ \\
\vspace{0.05cm}
$B_{c}\to D^{*0}\kappa^{+}(800)$&A&$55.2_{-9.4-7.4-1.8}^{+10.6+21.6+1.8}$&$0.10_{-0.25-0.17-0.01}^{+0.25+0.16+0.01}$ \\
\vspace{0.05cm}
$B_c\to D_s^{*} a_0^0(980)$&C&$0.06_{-0.01-0.01-0.00}^{+0.02+0.01+0.00}$&$-2.84_{-0.12-1.05-0.15}^{+0.29+4.48+0.20}$\\
\vspace{0.05cm}
$B_c\to D_s^{*} \sigma(f_n)$&P&$2.50_{-0.64-0.81-0.07}^{+0.72+1.68+0.08}$&$1.65_{-0.14-0.57-0.18}^{+0.13+0.33+0.16}$\\
\vspace{0.05cm}
$B_c\to D_s^{*}\sigma(f_s)$&A&$52.5_{-7.8-24.0-1.8}^{+7.8+8.8+1.8}$&0.0\\
\vspace{0.05cm}
$B_c\to D_s^{*}f_0(980)(f_n)$ &P&$2.50_{-0.64-0.81-0.07}^{+0.72+1.68+0.08}$&$1.65_{-0.13-0.57-0.18}^{+0.13+0.33+0.16}$\\
\vspace{0.05cm}
$B_{c}\to D_s^{*}f_{0}(980)(f_s)$&A&$108_{-14-50-4}^{+15+21+4}$&0.0\\
\vspace{0.05cm}
$B_{c}\to D_s^{*}\bar{\kappa}^0(800)$&A&$3.07_{-0.47-0.97-0.18}^{+0.47+1.69+0.17}$&$8.47_{-0.37-2.74-0.06}^{+0.51+2.63+0.05}$\\
 \hline\hline
\end{tabular}\label{T3}
\end{table}
 \begin{table}[!ht]
\centering
\caption{The CP-averaged branching fractions and CP asymmetries of $B_{c}\to D^* S (a_0(1450), K_0^*(1430), f_0(1370)$, and $f_0(1500))$  calculated in the PQCD approach in S1 and S2, respectively.}
\begin{tabular}[t]{l!{\;}c!{\;}c!{\;}c!{\;}c}
\hline\hline
  \multirow{2}{*}{Decay Modes} & \multirow{2}{*}{Class} & \multirow{2}{*}{BFs($10^{-6}$)} &\multirow{2}{*}{$A_{CP}(\%)$}&\multirow{2}{*}{Scenario}\\
  &&&&\\
 \hline
 \vspace{0.05cm}
$B_c\to D^{*+}a_{0}^0(1450)$&A&$2.18_{-0.59-0.90-0.48}^{+0.68+0.93+0.57}$&$-45.5_{-19.9-7.2-5.8}^{+18.4+15.1+4.9}$&$S_1$\\
\vspace{0.05cm}
&&$5.51_{-1.26-2.38-0.56}^{+1.42+2.81+0.62}$&$-43.7_{-10.4-6.1-0.6}^{+9.8+4.2+0.6}$&$S_2$\\
\vspace{0.05cm}
$B_c\to D^{*0}a_{0}^+(1450)$ &A&$3.31_{-0.96-1.27-0.46}^{+0.89+1.62+0.53}$&$-35.3_{-11.7-2.2-1.6}^{+14.3+22.1+1.9}$&$S_1$\\
\vspace{0.05cm}
&&$10.7_{-2.3-4.1-0.8}^{+2.8+5.1+0.8}$&$-16.5_{-4.6-3.5-0.4}^{+3.8+1.2+0.7}$&$S_2$\\
\vspace{0.05cm}
$B_c\to D^{*+}f_0(1370)$ &A&$2.55_{-0.70-0.51-0.11}^{+0.90+0.54+0.07}$&$0.12_{-5.58-2.32-0.20}^{+5.96+8.31+1.01}$&$S_1$\\
\vspace{0.05cm}
&&$4.26_{-0.82-1.60-0.11}^{+1.00+1.76+0.08}$&$24.2_{-6.6-1.5-2.6}^{+6.6+1.6+2.2}$&$S_2$\\
\vspace{0.05cm}
$B_c\to D^{*+}f_0(1500)$&A &$1.02_{-0.26-0.30-0.04}^{+0.27+0.42+0.02}$&$-4.03_{-12.48-6.81-0.13}^{+14.78+14.00+0.12}$&$S_1$\\
\vspace{0.05cm}
&&$2.35_{-0.47-1.08-0.07}^{+0.53+0.89+0.06}$&$31.3_{-7.8-1.2-2.8}^{+8.0+2.5+2.3}$&$S_2$\\
\vspace{0.05cm}
$B_c\to D^{*+}K_0^{*0}(1430)$ &A&$63.3_{-13.8-22.6-2.0}^{+14.6+23.7+2.0}$&0.0&$S_1$\\
\vspace{0.05cm}
&&$188_{-50-91-5}^{+51+93+4}$&0.0\&$S_2$\\
\vspace{0.05cm}
$B_c\to D^{*0}K_{0}^{*+}(1430)$&A&$61.7_{-12.6-13.3-1.6}^{+14.7+20.3+1.6}$&$-0.25_{-0.41-0.20-0.01}^{+0.41+0.33+0.02}$ &$S_1$\\
\vspace{0.05cm}
&&$192_{-46-83-4}^{+51+90+5}$&$-0.07_{-0.21-0.11-0.01}^{+0.21+0.13+0.01}$&$S_2$\\
\vspace{0.05cm}
$B_{c}\to D_s^{*}a_0^0(1450)$ &C&$0.14_{-0.05-0.04-0.01}^{+0.04+0.05+0..01}$&$-1.44_{-0.47-1.23-0.08}^{+0.32+1.92+0.11}$&$S_1$\\
\vspace{0.05cm}
&&$0.05_{-0.02-0.01-0.00}^{+0.02+0.01+0.00}$&$-0.15_{-0.87-1.06-0.01}^{+0.56+0.55+0.01}$&$S_2$\\
\vspace{0.05cm}
$B_{c}\to D_s^{*}f_0(1370)$&A&$34.2_{-9.2-10.8-0.1}^{+10.6+12.2+0.1}$&$-0.89_{-0.45-0.38-0.17}^{+0.55+0.20+0.16}$&$S_1$\\
\vspace{0.05cm}
&&$72.4_{-14.9-30.7-1.3}^{+17.1+35.1+1.3}$&$-2.09_{-0.52-0.32-0.18}^{+0.49+0.18+0.22}$&$S_2$\\
\vspace{0.05cm}
$B_c\to D_s^{*} f_0(1500)$&A&$28.5_{-6.8-7.9-0.8}^{+8.8+8.6+0.7}$&$0.95_{-0.80-0.11-0.15}^{+0.70+0.29+0.16}$&$S_1$\\
\vspace{0.05cm}
&&$186_{-42-82-5}^{+55+94+5}$&$1.01_{-0.25-0.10-0.04}^{+0.29+0.18+0.03}$&$S_2$\\
\vspace{0.05cm}
$B_c\to D_s^{*} \bar{K}_0^{*0}(1430)$&A&$4.87_{-0.97-1.34-0.28}^{+1.15+1.90+0.30}$&$-4.89_{-1.03-1.05-0.02}^{+0.80+0.68+0.03}$&$S_1$\\
\vspace{0.05cm}
&&$11.3_{-2.7-5.2-0.7}^{+2.9+6.3+0.6}$&$4.22_{-2.10-1.91-0.02}^{+1.09+0.16+0.02}$&$S_2$\\
\hline\hline
\end{tabular}\label{T4}
\end{table}
 \begin{table}[!htb]
\centering
 \caption{The calculated branching fractions of $B_{c}\to D^{(*)} f_0(980)$ and $\sigma$ with the mixing in the PQCD approach (unit:$10^{-6}$).}
\begin{tabular}[t]{l!{\;\;\;\;}c!{\;\;\;\;\;\;\;}c!{\;\;\;\;\;}c!{\;\;\;\;}c}
\hline\hline
  \multirow{2}{*}{Decay Modes} & \multicolumn{2}{c}{$[25^{\circ}, 40^{\circ}]$} & \multicolumn{2}{c}{$[140^{\circ}, 165^{\circ}]$} \\
  &BF($10^{-6}$)&$A_{CP}(\%)$&BF($10^{-6}$)&$A_{CP}(\%)$\\
 \hline
 \vspace{0.05cm}
$B_{c}\to D^{+}\sigma$&$2.82\sim2.04$&$-29.6\sim-26.3$&$1.95\sim3.14$&$-41.1\sim-36.1$\\
\vspace{0.05cm}
$B_{c}\to D^{+}f_0(980)$ &$0.94\sim2.04$&$-23.1\sim-26.7$&$1.77\sim0.25$&$-37.3\sim-49.9$\\
\vspace{0.05cm}
$B_c\to D_s\sigma$&$19.9\sim50.3$ &$-3.84\sim-1.96$&$63.9\sim13.4$&$1.57\sim3.82$\\
\vspace{0.05cm}
$B_c\to D_{s}f_0(980)$ &$160\sim117$&$0.60\sim1.06$&$104\sim172$&$-1.19\sim-0.37$\\
\vspace{0.05cm}
$B_c\to D^{*+}\sigma$ &$1.29\sim0.90$&$76.6\sim80.2$&$1.05\sim1.57$&$62.0\sim68.6$\\
\vspace{0.05cm}
$B_c \to D^{*+}f_0(980)$&$0.57\sim1.25$&$38.2\sim47.6$&$1.21\sim0.22$&$67.7\sim76.4$\\
\vspace{0.05cm}
$B_{c}\to D_s^{*}\sigma$ &$10.5\sim22.0$&$8.98\sim5.44$&$24.3\sim6.44$&$-4.72\sim-8.65$\\
\vspace{0.05cm}
$B_{c}\to D_s^{*}f_0(980)$&$89.8\sim65.1$&$-1.47\sim-2.59$&$64.0\sim101$&$2.68\sim0.86$\\
\hline\hline
\end{tabular}\label{T5}
\end{table}

As we know, the quark components and physical properties of the $f_0(980)$ and $\sigma$ are long-standing puzzles in particle physics. Although they are favored to be four-quark states, we here only assume the $f_0(980)$ and $\sigma$ to be $n\bar n$ and $s\bar{s}$ bound states with a mixing, because in four-quark scenario their wave functions and decay constants are still absent till now. Besides many measurements about the charmless $B$ decays involving a scalar meson, the LHCb collaboration has also reported their first measurements of the charmed $B$ decays with a scalar $B(B_s) \to \overline{D} \sigma$ and $\overline{D} f_0(980)$ decays \cite{lhcb} in the end of 2015. Although we have large amounts of data, the mixing angle $\theta$ cannot be constrained stringently due to the large uncertainties \cite{angle}. In this work, under 2-quark model, for the sake of convenience, we presented individually the branching fractions of the $B_c \to D^{(*)} \sigma/f_0(980)$ decays under the pure  $n\bar n$ and $s\bar{s}$ components. Once the 2-quark model were confirmed and the mixing angle were fixed by other experiments, the branching fractions can be directly obtained from our predictions. As aforementioned in Sec.II, we also presented in Table.\ref{T5} the branching fractions with mixing patterns by adopting two typical ranges, $[25^{\circ}, 40^{\circ}]$ and $[140^{\circ}, 165^{\circ}]$, where only the central values are quoted. As for $f_0(1370)$ and $f_0(1500)$,  after neglecting the negligible glueball contents,  the mixing form can be simplified as eq.~(\ref{eq:f02}). It is noted that in Tables \ref{T2} and \ref{T4}, the presented branching fractions are results with the mixing patterns.

As stated in ref.\cite{hejb}, the LHC experiment can produce about $10^9$ $B_c$ events every year. In ref.\cite{su3}, it has been estimated that the charmless $B_c$ decays with a branching fraction at the level $10^{-6}$ yield a few events per year at LHCb. Because the selection criteria and the trigger efficiencies are very different for each decay mode, in order to roughly estimate the expected sensitivity for the considered $B_c \to D S$ decays, the quantitative analysis based on the numerical results is necessary. Based on our predictions, we believe that some   $B_c \to D S$ decays with large decay rates will be detected in the experiments, such as the LHCb and CMS. Taking the decay $B_c \to D^+ K_0^{*0}(1430)$ as an example, the branching fraction is predicted to be about $1.9\times 10^{-4}$ in S1. In the experimental side, we in particular use the charged final states to reconstruct the $D^+$ and $K_0^{*0}(1430)$, the branching fractions of which are $BF(D^+ \to K^-\pi^+\pi^-)\simeq 10\%$ and $BF(K_0^{*0} \to K^+ \pi^-)\simeq 45\%$\cite{pdg}. According to ref.\cite{rcd}, if the total efficiency is assumed to be $1\%$, about 200 events per year can be expected in LHCb experiment. Since the branching fraction of $B_c \to D^+ K_0^{*0}(1430)$ in S2 is a bit larger than S1, we can expect more events to be detected. As for $B_c \to D^{*+} K_0^{*0}(1430)$ involving a vector charmed meson, the situation is similar, as the vector $D^{*}$ meson decays to $D$ meson with the rate close 100\%. Based on our predictions, the decays with the branching fractions ranging in $[10^{-5}, 10^{-4}]$ are expected to be measured in the near future.

On the basis of the numerical results we obtained, we give some discussions as follow:
\begin{itemize}
  \item For the decays with a scalar emitted, the contribution $\mathcal{A}_{ef}^{(*)LL}$ will vanish or be suppressed, because the neutral scalar meson can not be produced through local $(V\pm A)$ current, or the vector decay constants of the charged scalar mesons are highly suppressed by the tiny mass difference between the two running current quark masses. Since the factorizable emission diagrams are forbidden, the $B_c\to D_s^{(*)} a_0^0(980/1450)$ decays are only induced by the nonfactorizable emission diagrams $(C)$, therefore  these modes have tiny branching fractions. Generally, in contrast to the emission contributions, the annihilation type contributions are power suppressed in the charmless $B_{u,d,s}$  decays. However, in $B_c$ decays, the annihilation type contributions play the major contribution because of large enhancement by the Wilson coefficient $a_1$ and the CKM matrix elements $V_{cs(d)}$. In fact, this situation is  similar to the $B_c \to D^{(*)} T$ decays in ref.\cite{bcpqcd3} with $T$  denoting a  tensor meson. As a result, most of the considered decays are dominated by the $W$ annihilation $(A)$ type contributions, as classified in the tables. Specially, the $B_c \to D_{(s)}^{(*)+} \sigma /f_0(980)$ and $B_c \to D_{(s)}^{(*)+} f_0(1370/1500)$ decays are also dominated by the annihilations, though the final scalars are mixing between $n \bar n$ and $s\bar s$.
  \item Inevitably, there are large theoretical uncertainties in the numerical calculations, especially the scalar meson property are not well understood and the $B_c$ meson wave function are not accurate yet. In order to reduce the dependence of the input parameters, we thus define two ratios as
      \begin{eqnarray}
      &&\frac{BF(B_c\to D^{(*)0}a_0^+)}{BF(B_c\to D^{(*)+}a_0^0)}\sim 2,\\
      &&\frac{BF(B_c\to D^{(*)+}K_0^{*0})}{BF(B_c\to D^{(*)0}K_0^{*+})}\sim 1.
      \end{eqnarray}
      In ref.\cite{bcpqcd2}, it has been found that $Br(B_c \to D^{(*)0}\pi^+/\rho^+) \gg Br(B_c\to D^{(*)+}\pi^0/\rho^0)$, where the $ B_c \to D^{(*)0}\pi^+/\rho^+$ are dominated by the factorizable emission diagrams, while the color-suppressed modes $B_c\to D^{(*)+}\pi^0/\rho^0$ are dominated by the annihilation diagrams. The relation  $Br(B_c \to D^{(*)0}\pi^+/\rho^+) \gg Br(B_c\to D^{(*)+}\pi^0/\rho^0)$ means that in $B_c\to DP(V)$ the annihilation type contributions are suppressed, compared with the contributions of the factorizable emission diagrams. However, when the scalar is involved,  because both $ B_c\to D^{(*)0} a_0^+$ and $ B_c \to D^{(*)+} a_0^0$ are all dominated by the annihilation type contribution, the relation $BF(B_c\to D^{(*)0} a_0^+)$ $\sim 2 BF(B_c \to D^{(*)+} a_0^0)$  is understandable. Similar, we can also explain the relation of $BF(B_c\to D^{(*)+}K_0^{*0}) \approx BF(B_c\to D^{(*)0}K_0^{*+})$.

 \item In ref.\cite{bcpqcd2}, for the $B_c \to D^0 K^+$ and $B_c \to D^+ K^0$, besides the annihilations, they are also affected by the penguin operators. Compared with the contribution of annihilations, the penguin emission diagrams are sizable contribution but with a relative minus sign. So, their branching fractions are much smaller than those of the corresponding $B_c \to D^0 \kappa^+/K_0^{*+}(1430)$ and $B_c \to D^+ \kappa^0/K_0^{*0 }(1430)$ decays with the large annihilation contribution alone, because the emission contributions are highly suppressed. We note the magnitudes of  $B_c \to D^{(*)0} \kappa /K_0^{*}(1430)$ are at ${\cal O}(10^{-4})$, which is measurable in the LHCb. The measurement of such decays will afford a few hints for studying the annihilation mechanisms in $B$ physics.

  \item As expected, the branching fractions of the four ``C" type decays are much smaller than these ``A" type decays, and the contribution from penguin operators is negligible. Although the four decays are also with the large Wilson coefficient ($C_2 \sim 1.0$), they are suppressed by the tiny CKM factor. For example, the CKM factor of $B_c^+ \to D_sa_0$ is $V_{ub}^*V_{us}$, and the one of $B_c^+ \to D^+ a_0$ is $V_{ub}^*V_{ud}$.

  \item From the Tables. \ref{T2} and. \ref{T4}, we find that for these ``A'' type decays the branching fractions in S2 are about 2-3 times larger than those in S1, except the $B_c \to D_{(s)}^{(*)+}f_0(1370)/f_0(1500)$. However,  for these ``C'' type $B_c \to D_s^{(*)}a_0^0(1450)$ decays, the branching fractions in $S_2$ are much smaller than those in $S_1$, which illustrates that the contribution of the annihilation type diagrams in $S_2$ is much large than in $S_1$.  Somewhat differently, the contribution of hard-scattering emission diagrams in $S_2$ is much smaller than that in $S_1$. This phenomena is due to the different signs of the decay constants under different scenarios.

  \item We now discuss the decay modes involving the $f_0(1370, 1500)$ that are mixing states of $n\bar n$ and $s\bar s$. In $S_1$, the inference between the $n\bar n$ component and the $s\bar s$ component is destructive for $B_c \to D^+ f_0(1370)$ while constructive for $B_c \to D^+ f_0(1500)$. Therefore, the branching fraction of the $B_c \to D^+ f_0(1500)$ is about same as the $B_c \to D^+ f_0(1370)$, although the $B_c \to D^+ f_0(1500)$ is suppressed by the mixing coefficient with respect to the $B_c \to D^+ f_0(1370)$. For the $B_c \to D^{*+} f_0(1370)/f_0(1500)$ decays, the reverse applies. The inference is constructive (destructive) for $B_c \to D^+f_0(1500)$ ($B_c \to D^+f_0(1370)$), because the wave functions of $D$ and $D^*$ have different signs, as shown in Eqs.(\ref{DWF1}) and (\ref{DWF2}). As a result, the branching fraction of $B_c \to D^{*+} f_0(1500)$ is much smaller than that of $B_c \to D^{*+} f_0(1370)$. Similarly, the interference can also explain the relations $BF(B_c \to D_s f_0(1370)(S_1))\ll BF(B_c \to D_s f_0(1500)(S_1))$ and $BF(B_c \to D_s^* f_0(1370)(S_1))\sim BF(B_c \to D_s^* f_0(1500)(S_1))$. In $S_2$,  because the contributions from $ s \bar s$ (``C'' type) is negligible, the interference between $n\bar n$ and $s\bar s$ in $B_c \to D^{(*)+}f_0(1370)/f_0(1500)$ decays is weak, so that we could obtain
       \begin{eqnarray}
          \frac{BF(B_c \to D^+ f_0(1370))}{BF(B_c \to D^+ f_0(1500))} \sim \frac{BF(B_c \to D^{*+} f_0(1370))}{BF(B_c \to D^{*+} f_0(1500))} \sim 2.
       \end{eqnarray}
      Similarly, we have
       \begin{eqnarray}
           \frac{BF(B_c \to D_s f_0(1500))}{BF(B_c \to D_sf_0(1370))} \sim \frac{BF(B_c \to D_s^{*} f_0(1500))}{BF(B_c \to D_s^{*} f_0(1370))} \sim \frac{3}{2}.
       \end{eqnarray}
  \item  From the tables, it is apparent that the direct CP asymmetries are very small, since the contributions from penguin operators are much smaller than to those from the tree operators, except the $B_c \to D^{(*)+}a_0^0(980/1450)$ and $B_c \to D^{(*)+}\sigma/f_0(980)$ decays. For the $B_c \to D^{(*)+}a_0^0(980/1450)$, the contribution from tree operators is suppressed by the cancellation between the nonfactorizable emission diagrams and the annihilation type diagrams, so that the interference between the contributions from the tree operators and those from the penguin operators are sizable and the direct CP asymmetries become large. For the $B_c \to D^{(*)+} \sigma/f_0(980)$ decays, the contributions from the $f_s$ component are small, because the factorizable emission diagrams are forbidden and the nonfactorizable contributions are suppressed by the CKM matrix elements. That is reason why these decays are dominated by the $f_n$ component. For the $f_n$ component, the contribution from penguin operators is comparable to the one from tree operators, because the latter contribution becomes small due to the cancellation  between emission and  annihilation diagrams. So the $B_c \to D^{(*)+} \sigma/f_0(980)$ decays have large direct CP asymmetries in 2-quark picture. Unfortunately, these CP asymmetries cannot be measured in the current LHCb experiment, because their branching fractions  are too small. We also note that, the CP asymmetries of the $B_c \to D^{*+} f0(1370)/f_0(1500)$ decays are dependent on the scenarios heavily, which might be useful for identifying different scenario when the experiments are available in the near future.
\end{itemize}

\section{Conclusion}\label{sec:5}
In this work, within the PQCD approach, we studied the branching fractions and CP asymmetries of 40 $B_c \to DS$ decays where the scalar mesons are involved. For the $B_c$ meson, since it is the only heavy meson consisting of two heavy quarks with different flavor, its wave function are not well defined, we here adopted the $\delta$-function. For the scalars, because quark components of them have not been confirmed, two different scenarios have been discussed. It is worth noting that the nonperturbative parameters and the corrections from higher order and higher power are beyond the scope of this work and not included in this work, which can be left for the future work. After the calculation, we find several branching fractions are in the range of  $[10^{-5}, 10^{-4}]$, some of which could be measured in the LHCb experiment, and other decays with smaller fractions might be measured in the high-energy colliders. Furthermore, we also note that some decays have large CP asymmetries, but they are unmeasurable currently due to small branching fractions.

\section*{Acknowledgment}
This work was supported in part by the National Science Foundation of China under the Grant Nos.~11575151, 11705159, 11235005, 11765012, 11447032, and by the Natural Science Foundation of Shandong province (ZR2014AQ013 and ZR2016JL001), and also supported by the Research Fund of Jiangsu Normal University under Grant  No. HB2016004. YL is grateful to the institute of high energy physics (IHEP) for hospitality where this work is initiated.
\begin{appendix}
\section{Related Hard Functions}

In this appendix, we summarize the functions that appear in the analytic formulas in the Section \ref{sec:3}.
Firstly, we present the auxiliary functions as
\begin{eqnarray}
&&F_0^2=m^2_{B_c}(x_1-r_D^2)(1-x_3),\\
&&F_a^2=m^2_{B_c}(r_b^2- x_3(1-r_D^2)),\\
&&F_b^2=m_{B_c}^2(x_1-r_D^2);\\
&&F_c^2=m^2_{B_c}(x_3-1)((1-r_D^2)(1-x_2)-(x_1-r_D^2)),\\
&&F_d^2=m^2_{B_c}(x_3-1)((1-r_D^2)x_2-(x_1-r_D^2)),\\
&&E_0^2=m^2_{B_c}(x_{2}x_{3}(1-r_{D}^{2})),\\
&&F_e^2=m^2_{B_c}(x_3(1-r_D^2)),\\
&&F_f^2=m^2_{B_c}(((1-r_D^{2})x_2+r_D^2)-r_c^2),\\
&&F_g^2=m^2_{B_c}(r_b^2-(1-x_3)(1-x_1-x_2(1-r_D^2))),\\
&&F_h^2=m^2_{B_c}(r_c^2+x_3(x_1-x_2(1-r_D^{2}))).
\end{eqnarray}
The hard scales $t_i$ can be determined by
\begin{eqnarray}
t_{a}=\max\Big\{\sqrt{|F_0^2|},\sqrt{|F_a^2|}, 1/b_{1},1/b_{3}\Big\},\\
t_{b}=\max\Big\{\sqrt{|F_0^2|},\sqrt{|F_b^2|}, 1/b_{1},1/b_{3}\Big\},\\
t_{c}=\max\Big\{\sqrt{|F_0^2|},\sqrt{|F_c^2|}, 1/b_{1},1/b_{2}\Big\},\\
t_{d}=\max\Big\{\sqrt{|F_0^2|},\sqrt{|F_d^2|}, 1/b_{1},1/b_{2}\Big\},\\
t_{e}=\max\Big\{\sqrt{|E_0^2|},\sqrt{|F_e^2|}, 1/b_{2},1/b_{3}\Big\},\\
t_{f}=\max\Big\{\sqrt{|E_0^2|},\sqrt{|F_f^2|}, 1/b_{2},1/b_{3}\Big\},\\
t_{g}=\max\Big\{\sqrt{|E_0^2|},\sqrt{|F_g^2|}, 1/b_{1},1/b_{2}\Big\},\\
t_{h}=\max\Big\{\sqrt{|E_0^2|},\sqrt{|F_h^2|}, 1/b_{1},1/b_{2}\Big\}.
\end{eqnarray}
The hard functions are written as
\begin{multline}
 h_{a}=K_0(\sqrt{|F_0^2|} b_1)\\
\left\{\begin{array}{ll}
\theta(b_1-b_3)I_0( \sqrt{|F_a^2|} b_3)K_0( \sqrt{|F_a^2|} b_1)+\theta(b_3-b_1)I_0(m_{B_c}\sqrt{|F_a^2|} b_1)K_0( \sqrt{|F_a^2|} b_3)&F_a^2>0\\
\left[\theta(b_1-b_3)J_0( \sqrt{|F_a^2|} b_3)H_0^{(1)}( \sqrt{|F_a^2|} b_1) +\theta(b_3-b_1)J_0( \sqrt{|F_a^2|} b_1)H^{(1)}_0( \sqrt{|F_a^2|} b_3)\right]&F_a^2<0
\end{array}\right.,
\end{multline}
\begin{multline}
 h_{b}=K_0(\sqrt{|F_0^2|} b_3)\\
\left\{\begin{array}{ll}
\theta(b_1-b_3)I_0(\sqrt{|F_b^2|}  b_3)K_0\sqrt{|F_b^2|} b_1)+\theta(b_3-b_1)I_0(\sqrt{|F_b^2|} b_1)K_0(\sqrt{|F_b^2|}  b_3)&F_b^2>0\\
\left[\theta(b_1-b_3)J_0(\sqrt{|F_b^2|}  b_3)H_0^{(1)}(\sqrt{|F_b^2|}  b_1) +\theta(b_3-b_1)J_0(\sqrt{|F_b^2|} b_1)H^{(1)}_0(\sqrt{|F_b^2|}  b_3)\right]&F_b^2<0
\end{array}\right.,
\end{multline}

\begin{multline}
h_{c}=\left[\theta(b_{2}-b_{1})K_{0}(\sqrt{|F_0^2|}b_2)I_{0}(\sqrt{|F_0^2|}b_1) +\theta(b_{1}-b_{2})K_{0}(\sqrt{|F_0^2|}b_1)I_{0}(\sqrt{|F_0^2|}b_2)\right] \\
 \cdot \left\{\begin{array}{ll}
\frac{i\pi}{2}H_{0}^{(1)}\left(\sqrt{|F_c^2|}b_{2}\right),& \;\;F_c^2<0;\\
K_{0}\left(\sqrt{|F_c^2|}b_{2}\right),&\;\;F_c^2>0,
\end{array}\right.
\end{multline}

\begin{multline}
h_{d}=\left[\theta(b_{2}-b_{1})K_{0}(\sqrt{|F_0^2|}b_2)I_{0}(\sqrt{|F_0^2|}b_1) +\theta(b_{1}-b_{2})K_{0}(\sqrt{|F_0^2|}b_1)I_{0}(\sqrt{|F_0^2|}b_2)\right] \\
 \cdot \left\{\begin{array}{ll}
\frac{i\pi}{2}H_{0}^{(1)}\left(\sqrt{|F_d^2|}b_{2}\right),& \;\;F_d^2<0;\\
K_{0}\left(\sqrt{|F_d^2|}b_{2}\right),&\;\;F_d^2>0,
\end{array}\right.
\end{multline}

\begin{multline}
h_{e}=(\frac{i\pi}{2})^{2}H_{0}^{(1)}\left(\sqrt{|E_0^2|}b_{2}\right) \\
 \left[\theta(b_{2}-b_{3})H_{0}^{(1)}\left(\sqrt{|F_{e}^{2}|} b_{2}\right)J_{0}\left(\sqrt{|F_{e}^{2}|} b_{3}\right)+\theta(b_{3}-b_{2})H_{0}^{(1)}\left(\sqrt{|F_{e}^{2}|} b_{3}\right)J_{0}\left(\sqrt{|F_{e}^{2}|}b_{2}\right)\right]\cdot
S_{t}(x_{3}),
\end{multline}

\begin{multline}
h_{f}=(\frac{i\pi}{2})^{2}H_{0}^{(1)}\left(\sqrt{|E_0^2|}b_{2}\right) \\
 \left[\theta(b_{2}-b_{3})H_{0}^{(1)}\left(\sqrt{|F_{f}^{2}|} b_{2}\right)J_{0}\left(\sqrt{|F_{f}^{2}|} b_{3}\right)+\theta(b_{3}-b_{2})H_{0}^{(1)}\left(\sqrt{|F_{f}^{2}|} b_{3}\right)J_{0}\left(\sqrt{|F_{f}^{2}|}b_{2}\right)\right]\cdot
S_{t}(x_{3}),
\end{multline}

\begin{multline}
h_{g}=\frac{i\pi}{2}\left[\theta(b_{1}-b_{2})H_{0}^{(1)}\left(\sqrt{|E_0^2|}b_{1}\right)J_{0}\left(\sqrt{|E_0^2|}b_{2}\right) +\theta(b_{2}-b_{1})H_{0}^{(1)}\left(\sqrt{|E_0^2|}b_{2}\right)J_{0}\left(\sqrt{|E_0^2|}b_{1}\right)\right] \\
 \times \left\{\begin{array}{ll}
\frac{i\pi}{2}H_{0}^{(1)}\left(\sqrt{|F_g^2|}b_{1}\right),&
F_{g}^{2}<0,\\
K_{0}\left(\sqrt{|F_g^2|}b_{1}\right),& F_{g}^{2}>0,
\end{array}\right.
\end{multline}

\begin{multline}
h_{h}=\frac{i\pi}{2}\left[\theta(b_{1}-b_{2})H_{0}^{(1)}\left(\sqrt{|E_0^2|}b_{1}\right)J_{0}\left(\sqrt{|E_0^2|}b_{2}\right) +\theta(b_{2}-b_{1})H_{0}^{(1)}\left(\sqrt{|E_0^2|}b_{2}\right)J_{0}\left(\sqrt{|E_0^2|}b_{1}\right)\right] \\
 \times \left\{\begin{array}{ll}
\frac{i\pi}{2}H_{0}^{(1)}\left(\sqrt{|F_h^2|}b_{1}\right),&
F_{h}^{2}<0,\\
K_{0}\left(\sqrt{|F_h^2|}b_{1}\right),& F_{h}^{2}>0.
\end{array}\right.
\end{multline}

The $S_{t}(x)$ is the jet function  from the threshold resummation,  which can be written as \cite{jet}
\begin{eqnarray}
S_{t}(x)\,=\,\frac{2^{1+2c}\Gamma(3/2+c)}{\sqrt{\pi}\Gamma(1+c)}[x(1-x)]^{c},
\end{eqnarray}
with $c=0.3$. The evolution functions $E_{i}$ and $E_{enf}(t_{b})$ in the analytic formulas are given by
\begin{eqnarray}
&&E_{ef}(t)=\alpha_{s}(t)\exp[-S_{B_c}(t)-S_{D}(t)];\\
&&E_{enf}(t)=\alpha_{s}(t)\exp[-S_{B_c}(t)-S_{D}(t)-S_{S}(t)]| _{b_{1}=b_{3}};\\
&&E_{af}(t)=\alpha_{s}(t)\exp[-S_{D}(t)-S_{S}(t)]|;\\
&&E_{anf}(t)=\alpha_{s}(t)\exp[-S_{B_c}(t)-S_{D}(t)-S_{S}(t)]|_{b_{2}=b_{3}}.
\end{eqnarray}
The Sudakov exponents are defined as
\begin{eqnarray}
S_{B_c}(t)\,=\,s\left(x_{1}\frac{m_{B_c}}{\sqrt{2}},b_{1}\right)\,+\,\frac{5}{3}\int_{1/b_{1}}^{t}\frac{d\bar{\mu}}{\bar{\mu}}\gamma_{q}(\alpha_{s}(\bar{\mu})),
\end{eqnarray}
\begin{eqnarray}
S_{D}(t)\,=\,s\left(x_{3}\frac{m_{B_c}}{\sqrt{2}},b_{3}\right)
\,+\,2\int_{1/{b_3}}^{t}\frac{d\bar{\mu}}{\bar{\mu}}\gamma_{q}(\alpha_{s}(\bar{\mu})),
\end{eqnarray}
\begin{eqnarray}
S_{S}(t)\,=\,s\left(x_{2}(1-r_D^2)\frac{m_{B_c}}{\sqrt{2}},b_{2}\right)\,+\,s\left((1-x_{2})(1-r_D^2)\frac{m_{B_{c}}}{\sqrt{2}},b_{3}\right)
\,+\,2\int_{1/{b_2}}^{t}\frac{d\bar{\mu}}{\bar{\mu}}\gamma_{q}(\alpha_{s}(\bar{\mu})),
\end{eqnarray}
where the $s(Q,b)$ can be found in ref.\cite{kt1}.
\end{appendix}

\end{document}